\documentclass[twocolumn,dvipsnames]{aastex631}

\usepackage{apjfonts}
\DeclareSymbolFont{cmletters}{OML}{cmm}{m}{it}
\DeclareMathSymbol{v}{\mathalpha}{cmletters}{"76} 
\usepackage{hyperref}

\usepackage{amssymb}
\usepackage{amsmath}

\usepackage{xspace}

\newcommand{\chandra}{{\it Chandra}\xspace}

\newcommand{\eqb}{\begin{eqnarray}}
\newcommand{\eqe}{\end{eqnarray}}

\newcommand{\mdot}{\ensuremath{\dot M}\xspace}
\newcommand{\edot}{\ensuremath{\dot E}\xspace}

\newcommand{\mdotb}{\ensuremath{\dot M_{\rm B}}\xspace}
\newcommand{\rg}{\ensuremath{R_{\rm g}}\xspace}
\newcommand{\rb}{\ensuremath{R_{\rm B}}\xspace}
\newcommand{\rcirc}{\ensuremath{R_{\rm circ}}\xspace}
\newcommand{\BH}{{{\rm BH}}\xspace}
\newcommand{\SMBH}{{{\rm SMBH}}\xspace}
\newcommand{\MBH}{{M_{\rm BH}}\xspace}
\newcommand{\der}{{\rm d}}
\newcommand{\p}{{\partial}}

\newcommand\hammer{\texttt{H-AMR}\xspace}

\newcommand{\ore}[1]{\textcolor{magenta}{}}
\newcommand{\aris}[1]{\textcolor{red}{}}
\newcommand{\sasha}[1]{\textcolor{OliveGreen}{}}
\newcommand{\nick}[1]{\textcolor{green}{}}
\newcommand{\irina}[1]{\textcolor{orange}{}}
\newcommand{\lena}[1]{\textcolor{blue}{}}
\newcommand{\kc}[1]{\textcolor{MidnightBlue}{}}
\newcommand{\ian}[1]{\textcolor{purple}{}}

\shorttitle{X-shaped Jets from GRMHD simulations}
\shortauthors{Lalakos et al.}
\graphicspath{{./}{figures/}}
\usepackage{graphicx}

\begin{document}

\title{Bridging Bondi and Event Horizon Scales: 3D GRMHD Simulations Reveal X-Shaped Radio Galaxy Morphology}

\author{Aretaios Lalakos}
\email{aretaioslalakos2022@u.northwestern.edu}
\affiliation{Center for Interdisciplinary Exploration \& Research in Astrophysics (CIERA), Physics \& Astronomy, Northwestern University, Evanston, IL 60202, USA}

\author{Ore Gottlieb}
\affiliation{Center for Interdisciplinary Exploration \& Research in Astrophysics (CIERA), Physics \& Astronomy, Northwestern University, Evanston, IL 60202, USA}

\author{Nick Kaaz}
\affiliation{Center for Interdisciplinary Exploration \& Research in Astrophysics (CIERA), Physics \& Astronomy, Northwestern University, Evanston, IL 60202, USA}

\author{Koushik Chatterjee}
\affiliation{Black Hole Initiative at Harvard University, 20 Garden Street, Cambridge, MA 02138, USA}
l
\author{Matthew Liska}
\affiliation{Institute for Theory and Computation, Harvard University, 60 Garden Street, Cambridge, MA 02138, USA; John Harvard Distinguished Science and ITC Fellow}

\author{Ian M. Christie}
\noaffiliation

\author{Alexander Tchekhovskoy}
\affiliation{Center for Interdisciplinary Exploration \& Research in Astrophysics (CIERA), Physics \& Astronomy, Northwestern University, Evanston, IL 60202, USA}

\author{Irina  Zhuravleva}
\affiliation{Department of Astronomy and Astrophysics, The University of Chicago, Chicago, IL 60637, USA}

\author{Elena Nokhrina}
\affiliation{Moscow Institute of Physics and Technology, Dolgoprudny, Institutsky per., 9, Moscow region, 141700, Russia}



\begin{abstract}
X-shaped radio galaxies (XRGs) produce misaligned X-shaped jet pairs and make up $\lesssim10$\% of radio galaxies. XRGs are thought to emerge in galaxies featuring a binary supermassive black hole (\SMBH), \SMBH merger, or large-scale ambient medium asymmetry. We demonstrate that XRG morphology can naturally form without such special, preexisting conditions. Our 3D general-relativistic magnetohydrodynamic (GRMHD) simulation for the first time follows magnetized rotating gas from outside the \SMBH sphere of influence of radius \rb to the \SMBH of gravitational radius \rg, at the largest scale separation $\rb/\rg = 10^3$ to date. Initially, our axisymmetric system of constant-density hot gas contains weak vertical magnetic field and rotates in an equatorial plane of a rapidly spinning \SMBH. We seed the gas with small-scale $2$\%-level pressure perturbations. Infalling gas forms an accretion disk, and the \SMBH launches relativistically-magnetized collimated jets reaching well outside \rb. Under the pressure of the infalling gas, the jets intermittently turn on and off, erratically wobble, and inflate pairs of cavities in different directions, resembling an X-shaped jet morphology. Synthetic X-ray images reveal multiple pairs of jet-powered shocks and cavities. Large-scale magnetic flux accumulates on the \SMBH, becomes dynamically important, and leads to a magnetically arrested disk state. The \SMBH accretes at $2$\% of the Bondi rate ($\mdot\simeq2.4\times10^{-3}M_{\odot}\,{\rm yr}^{-1}$ for M87*), and launches twin jets at $\eta=150$\% efficiency. These jets are powerful enough ($P_{\rm jets}\simeq2\times10^{44}\,{\rm erg\,s}^{-1}$) to escape along the spin axis and end the short-lived jets state whose transient nature can account for the rarity of XRGs.
\end{abstract}



\section{Introduction} \label{sec:intro}
Modeling how the hot interstellar medium (ISM) or intercluster medium (ICM) reaches from the outskirts of a galaxy or cluster of galaxies down to the central super-massive black hole (\SMBH) remains an unsolved, grand challenge problem. Observationally, it is clear that some poorly constrained fraction of the gas makes it down all the way to the \SMBH, forms an accretion disk, and powers an active galactic nucleus (AGN). Such systems can produce copious radiation and mechanical outflows, broad winds and tightly collimated relativistic jets. The jets can propagate through the galaxy, generate shocks, inflate bubbles, and heat up the the surrounding medium \citep{2016MNRAS.458.2902Z,2020ApJ...889L...1L, martizzi2019simulations}. It's commonly accepted that jet launching near the event horizon is powered by extraction of black hole (\BH) rotational energy \citep{blanford1977}.
The jets initially align with the \BH spin axis \citep{2013Sci...339...49M}, but interactions with the disk outflows can change their direction and they follow the angular momentum of the accretion disk \citep{liska2018formation}. Thus, the jet direction is determined by the complex interplay between the properties of the \BH and its feeding.

A small but significant, $\sim 5-10\%$, fraction of observed radio galaxies displays an intriguing X-shaped morphology, suggesting the presence of two pairs of jets at different angles \citep{1984MNRAS.210..929L,2019ApJS..245...17Y}.
Such X-shaped radio galaxies (XRGs) have been hypothesized to emerge due to:
(i)~restarted activity of the central \BH producing jets with different orientation \citep{bruni2019discovery, bruni2020hard},
(ii)~jet deflection by the hot and oblique halo/ISM of the galaxy \citep{cotton2020hydrodynamical},
(iii)~the presence of a \SMBH binary, each component producing its own pair of jets \citep{2007MNRAS.374.1085L},
(iv)~reorientation of the \SMBH spin axis due to \SMBH coalescence \citep{2002Sci...297.1310M,roberts2015abundance}, and
(v)~precession of the jet axis \citep{parma1985high} due to e.g.
Lense-Thirring precession of a tilted accretion disk \citep{liska2018formation}. These might not be the only scenarios for forming the X-shaped morphology, and combinations of these are also plausible. 

The goal of this work is to investigate whether X-shaped jet morphology can emerge spontaneously, without the above special, preexisting conditions. 
To reach the \SMBH, the gas needs to make its way from the ISM/ICM, located outside the sphere of influence of the \SMBH, or the Bondi radius $\rb = G\MBH/c^2_\infty \sim 20\ {\rm pc} \times(\MBH/10^9M_\odot)$, at which the thermal pressure is high enough to counter the free-fall of the gas \citep{bondi1952}, down to the scale of the gravitational radius, $\rg = G\MBH/c^2 = 5\times10^{-5}\ {\rm pc}\times(\MBH/10^9M_\odot)$. Here, $\MBH$ is the \SMBH mass and $c_\infty$ is the sound speed in the ISM/ICM.
Galaxy simulations have been able to follow the gas flow down to $\sim 1$~kpc \citep[e.g.,][]{alcazar2015,alcazar2017} and even sub-pc scales in hyper-refined Lagrangian simulations \citep{2021ApJ...917...53A}, approaching and even reaching inside of the Bondi sphere. However, the event horizon is still orders of magnitude below the parsec scale, demanding dedicated numerical effort to bridge this gap in scale separation.

The simplest description of accretion is the spherically-symmetric analytic \citet{bondi1952}  model, which does not include any rotation. 
Bondi accretion with nonzero angular momentum has been simulated by multiple groups \citep[e.g.,][]{2003ApJ...582...69P,2003ApJ...592..767P,2012ApJ...744..185C,li2013,2015MNRAS.447.1565S,2017MNRAS.472.4327S,2019MNRAS.488.5162X,kaaz2019,palit2019time,waters2020},  often utilizing axisymmetric, non-relativistic, hydrodynamic simulations. Accretion from a realistic ISM will be necessarily magnetized and have at least some angular momentum support.
Thus, it is important to model both rotation and magnetic fields. Rotation breaks the symmetry of the problem, provides rotational support, and tends to enhance mass outflows, reducing the mass reaching the \BH. Magnetic fields are important, as they form magnetically-powered outflows that can inject energy into the ambient gas. It is also crucial to include general relativistic (GR) effects, in order to accurately represent the energy and momentum feedback by the central \SMBH via launching \BH-powered jets and disk-powered winds and properly account for the poorly understood effects the \BH has on the inner boundary condition of the flow. It is also critical to extend the studies to 3D, to model the essentially non-axisymmetric magnetized turbulence, the associated angular momentum transport in the accretion disk, and to account for the development of 3D magnetic kink instabilities in the jets \citep[see also][]{ressler2021magnetically,2022arXiv220111753K,jia2022observational}. \citet{tchekhovskoy2016} initiated relativistic jets by the magnetized rotation at the inner grid boundary of radius $\rb=0.1$ kpc and followed their propagation through the ISM over distances of tens of kpc for long durations, sufficient for the instabilities to develop. They found that the jets inflated cavities, morphologically similar to M87* and other low-luminosity AGN as seen, e.g., by \chandra \citep{forman2017}. \citet{barniol_duran2017} showed that the jets can undergo internal 3D kink instabilities and dissipation triggered by the change in the radial density profile at the Bondi radius \citep{russell2015}. 

Incorporating all of these effects, 3D GRMHD simulations are unique tools enabling self-consistent studies of gas accretion from the Bondi scale, including the formation of a turbulent accretion disk, the launching of relativistic jets, and the development of 3D magnetic kink instabilities that can dramatically affect the jet morphology and the state of the ambient gas.
Despite the success of GRMHD simulations, direct modeling of 3D accretion spanning the full range of $\log_{10}(\rb/\rg) \simeq 5{-}6$ orders of magnitude in distance (and $8{-}9$ orders of magnitude in time) between the \BH and Bondi scales still remains computationally prohibitive. Our approach is to reduce the scale separation to the maximum that is computationally feasible. 
While the scale separation is smaller than in reality, the more orders of magnitude it spans, the closer the flows can approach a self-similar regime that approximates nature. By varying the degree of scale separation, we can evaluate the effects our particular choices for the $\rb/\rg$ ratio have on the structure of the flow, and analytically extend the results to the values of $\rb/\rg$ ratio found in nature. 

In this Letter, we present the results of the first 3D GRMHD simulation that follows the accretion of an initially magnetized-rotating gas from a separation scale up to $\rb/\rg=10^3$, which in 3D GRMHD is the largest to date and close to the maximum possible value attainable with current computational resources when evolved over astrophysically-interesting times. Starting with the simplest, axisymmetric state of constant density, rotational profile of the gas, and vertical magnetic fields outside $\rb$, we evolve the system until the formation of an accretion disk, a pair of relativistic jets, and their interaction with the ambient gas.
In Sec.~\ref{sec:setup} we present the numerical setup and the physical parameters of our model, and in Sec.~\ref{sec:results} we present the results: the emergence of the X-shaped morphology (Sec.~\ref{sec:intermittent}), the jet shape (Sec.~\ref{sec:jet-shape}), and synthetic X-ray images (Sec.~\ref{sec:synthetic}). In Sec.~\ref{sec:discussion} we conclude. We use units of $G=M=c=1$.

\section{Setup} \label{sec:setup}

We carry out a 3D GRMHD simulation using our GPU-accelerated GRMHD code \hammer (\citealt{2019arXiv191210192L}, see \citealt{Porth:19} for comparisons with other current GRMHD codes). 
We employ spherical polar coordinates, $r$, $\theta$, $\varphi$, and choose a spherical polar grid that is uniform in $\log r$, $\theta$, and $\varphi$ variables. The grid spans the range of $(0.97\rg, 10^5\rg) \times (0,\pi) \times (0,2\pi)$ and has the resolution of $N_r\times N_{\theta} \times N_{\varphi}=192\times256\times128$ cells in the $r$-, $\theta$-, and $\varphi$-directions, respectively. We choose it such that the first 5 radial cells are inside the event horizon: this ensures that the \BH exterior is causally disconnected from the inner radial grid boundary. We use outflow boundary conditions in the radial, transmissive conditions in the polar, and periodic conditions in the azimuthal directions.
We adopt axisymmetric initial conditions described by the following physical parameters:
\\\textbf{(i) Density profile}: Outside the Bondi radius, we set the ambient density to a uniform value of $\rho_\infty=1$. Since the simulation includes neither the cooling (radiation) nor self-gravity effects, the results are scale-free in density: the simulation results can be freely rescaled for any value of $\rho_{\infty}$. Within the Bondi radius, we place an empty cavity. The reason for this is we do not want to impose an initial density profile that may influence the subsequent steady-state solution. 
\\\textbf{(ii) Bondi radius:} As we discuss in Sec.~\ref{sec:intro}, in order to make the simulations affordable computationally, we choose a smaller \rb\ than inferred from observations ($\rb \sim 10^{5-6}\rg$). Namely, we adopt a value of $\rb=10^3\rg$: this is the largest scale separation between \rb and \rg that has ever been achieved in 3D GRMHD that we are aware of, yet it still allows us to evolve the system for astrophysically-interesting times. When discussing large-scale jet propagation outside \rb, we
convert the simulated lengths and times to physical lengths and times by associating the Bondi scale and its light crossing time in simulation units, $\rb$ and $\rb/c = 10^3 \rg/c$, with those in the physical units for M87*, $\rb = 0.15$~kpc and $\rb/c = 0.15\ {\rm kpc}/c =  475\ {\rm yr}$, respectively \citep{russell2015}. Our choice of $\rb/\rg$ naturally sets the sound speed in the ambient gas, $c_s=(\rb/\rg)^{-1/2}c= 10^{-3/2}c\approx0.03c$. We adopt a non-relativistic ideal gas equation of state, $p_{\rm g}= (\gamma-1)u_g$, where $\gamma=5/3$ is the polytropic index for a monatomic gas, and $u_{\rm g}$ and $p_{\rm g}$ are the gas internal energy density and pressure as measured in the comoving frame of the gas.
\\\textbf{(iii) Circularization radius:} We assign a rotational profile to the gas, such that on each spherical shell of radius $r$, the gas undergoes solid-body rotation at the angular velocity, $\omega = l_0/r^2$, around the $z$-axis. Here, $l_0$ characterizes the specific angular momentum of the gas in the equatorial plane. We choose $l_0 = {\rm constant}$, such that gas specific angular momentum, $l \simeq l_0\sin^2\theta$, reaches its maximum value at the equatorial plane and smoothly drops down to zero near the poles \cite[see also][]{palit2019time}.\footnote{Here, $l = u_\varphi \equiv g_{\varphi\mu}u^\mu \approx r^2\sin^2\theta\,\der\varphi/\der t$ is the $\varphi$-component of the covariant four-velocity, a conserved quantity for test particles in the Kerr space-time (we evaluated the approximate equality above in the non-relativistic limit and flat space).}
The value of $l_0$ depends on the circularization radius, \rcirc, at which the angular momentum in the equatorial plane equals the local Keplerian value. We adopt 
\begin{equation}
    \rcirc=l_0^2/(G\MBH) = 30R_g, 
    \label{eq:rcirc}
\end{equation}
which divides approximately equally the scale separation between \rg and \rb and results in sub-Keplerian rotation at the Bondi radius, $l_0/l_{\rm K}(\rb) = (\rcirc/\rb)^{1/2}\approx 0.17$.
\\\textbf{(iv) Initial gas magnetization:} We adopt the initial magnetic field threading the gas to be asymptotically a homogeneous vertical magnetic field in the $z$-direction and initialize it by setting the covariant magnetic vector potential, which has only one non-zero component: $A_{\varphi}=(r^2-\rb^2) \sin^2\theta$ at $r\ge \rb$ and $0$ otherwise. This ensures that $A_\varphi$ and the magnetic field vanishes inside \rb and does not bias the formation and evolution of the accretion flow there. Physically, the covariant $\varphi$-component of the vector potential, $A_\varphi$, gives the poloidal (pointing in the $r$- and $\theta$-directions) magnetic flux enclosed by an axisymmetric ring, $(r,\theta)$, divided by $2\pi$. We characterize the strength of the magnetic field via the plasma-$\beta$ parameter, defined as the ratio of thermal to magnetic pressure, $\beta=p_{\rm g}/p_{\rm m}$, where $p_{\rm m}$ is the magnetic pressure. We normalize the magnetic field strength by choosing the characteristic value of plasma-$\beta$ to be high enough so that the initial accretion stage is gas pressure-dominated, $\min\beta = 100$.
\\\textbf{(v) BH spin:} We consider a rapidly spinning \BH, with high dimensionless spin, $a = 0.9375$, to give the jets the best chance to form and survive in their fight against the onslaught of the infalling gas: if the jets form, their power will be close to the maximum possible value for a given magnetic flux ($P_{\rm jets} \propto a^2$, \citealt{blanford1977}), enabling us to study the physics of their interaction with the infalling gas.

We add to the initial pressure random perturbations  (independent for each numerical cell)  at the level of
2\%: without the perturbations, the system would
maintain the exact axisymmetry and its evolution would be identical to
a 2D model. 
We then let the system evolve out to $t=2.3\times
10^5\rg/c = 230\rb/c = 0.1$~Myr and report the simulation results.

\section{Results} \label{sec:results}
\subsection{Natural Development of X-shaped Jet Morphology} \label{sec:intermittent}

\begin{figure}[!htp]
{\includegraphics[width=\columnwidth]{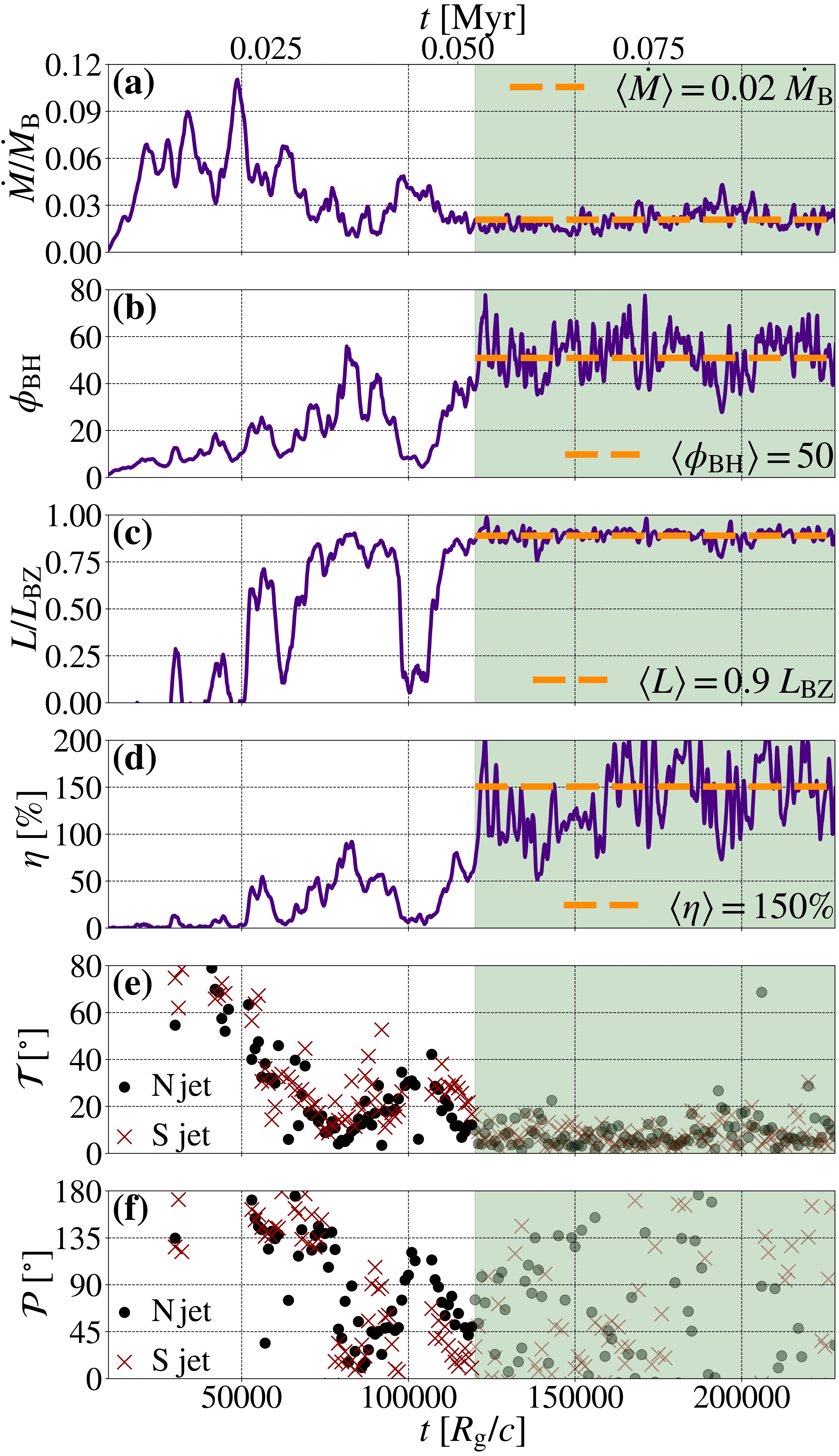}}
\caption{To survive the onslaught of the infalling gas, jets deflect towards the equatorial plane ($\mathcal T\sim 90^\circ$, panel e), before aligning with the $z$-axis ($\mathcal T\lesssim 10^\circ$). This jet reorientation naturally results in an X-shaped jet morphology (Fig.~\ref{fig:plots_grid}). The early period, $t\le 5\times10^4 \rg/c$, of high mass accretion rate $\mdot/\mdotb\gtrsim0.06$ (panel a) features low values of the dimensionless \BH magnetic flux $\phi_{\rm BH}\lesssim20$ (panel b), total-EM to BZ jet-power ratio $L/L_{\rm BZ}
\lesssim0.25$ (panel c), and jet energy efficiency $\eta
\lesssim10$\% (panel c). The infalling dense gas easily deflects the weak jets sideways, towards the equatorial plane, resulting in high jet tilt angle $\mathcal T\sim 90^\circ$ (panel e) and large variations, $\sim 180^\circ$, in precession $\mathcal P$ angle (panel f). At later times, $t \gtrsim 10^5 \rg/c$, the accumulation of large-scale vertical magnetic flux on the \BH leads to strong $\phi_{\rm BH}\sim 50$ (panel b) and a MAD state (highlighted in green; steady state averages shown with horizontal dashed orange lines): the powerful jets, $L/L_{\rm BZ}\sim 1$ (panel c) and $\eta\sim 150$\% (panel d), suppress $\mdot$ (panel a) and stably propagate along the $z$-axis ($\mathcal T\lesssim 10^\circ$, panel e), about which both the \BH and the ambient gas rotate, out to large distances (Figs.~\ref{fig:plots_grid} and \ref{fig:jet-radius}). During the MAD state, $\mdot$ saturates at $2$\% of $\mdotb$ (panel a), implying that only $2$\% of the Bondi accretion rate reaches the \BH and the rest $98$\% leaves as outflows.}
\label{fig:q_vs_time}
\end{figure}

The simulation starts\footnote{ A link with of movies is provided here: \href{https://www.youtube.com/playlist?list=PLTZ8Aqtsc3rGjgpnjqtR7tH6jsjtzemXJ}{youtube.com/movies}}
with a uniform gas distribution outside the Bondi radius, and a vacuum hole at $r < \rb$. The gravitational forces along with the pressure gradient at the interface of the hot gas and the empty cavity, at $r = \rb$, push the gas inward. Because the gas possesses a non-zero angular momentum (Sec.~\ref{sec:setup}), it undergoes what is nearly a free-fall until it hits the centrifugal barrier at $\rcirc = 30\, \rg$. At this point, the radial infall slows down, and inside \rcirc an accretion disk forms. Because on the way to the \BH the gas spends most of the time at the largest distances, we can estimate the time it takes for it to travel from \rb to the \BH as the free-fall time, $t_{\rm ff} = 2^{-1/2}(\rb/\rg)^{3/2}\rg/c \approx 2.2\times 10^4\rg/c$. As seen in Fig.~\ref{fig:q_vs_time}(a), around $t_{\rm ff}$ the mass accretion rate $\dot M$ reaches its first peak and oscillates thereafter. Here, $\dot{M} = -\iint  \rho u^r \der A$, where $\der A=\sqrt{-g}\der\theta \der\varphi$ and $g= 
\left|g_{\mu \nu}\right|$ is the determinant of the metric. Analogously, we define the energy accretion rate, $\dot{E}=\iint[(\rho + u_g + p_g + 2p_{\rm m})u^r u_t -b^r b_t/4\pi] \der A$, where $b^\mu$ is the fluid frame magnetic field 4-vector. We evaluate both $\dot M$ and $\dot E$ at $r = 5\rg$, to avoid potential contamination by the density floors near the horizon. Because in a steady state $\dot M$ and $\dot E$ are conserved and independent of radius, this does not affect time-average values and only slightly shifts the dependencies in time by $\Delta t
\lesssim 5\rg/c$, i.e., shorter than the sampling time of the simulation. 

We use the sign convention such that positive $\mdot$ and $\edot$ imply mass and energy entering the \BH. We also define the accretion efficiency as $\eta=(\dot{E}-\mdot)/\mdot$. The gas drags with it the magnetic flux, and as Fig.~\ref{fig:q_vs_time}(b) shows, this leads to an increase in the absolute magnetic flux on the \BH, $\Phi_\BH = 0.5\iint \left|B^r\right| \der A$, where the integration is over the area of the event horizon. We further normalize the magnetic flux by the time-smoothed mass accretion rate, $\phi_\BH \equiv \Phi_\BH/\left(\dot M \rg^2 c\right)^{1/2}$. 
We smooth all quantities in Fig.~\ref{fig:q_vs_time} over a timescale of $1000\rg/c$ using a 0th-order Savitzky-Golay filter, to make the plots more readable.

As the accretion disk forms, magnetized rotation of the \BH works to form highly-magnetized jets. 
The increasing magnetic flux on the \BH would ordinarily translate into magnetically-powered jets via the \citet[BZ, hereafter]{blanford1977} effect. However, the \BH is engulfed in the infalling gas from all directions, which suppresses the event horizon electromagnetic (EM) luminosity of the jets, $L$, well below the analytic BZ expectation, $L_{\rm BZ}$: Fig.~\ref{fig:q_vs_time}(c) shows that $L/L_{\rm BZ} \lesssim 0.25$ at $t\lesssim 5\times 10^4\rg/c$.
In order for the jets to successfully launch and avoid falling apart due to gas-jet interaction and the development of magnetic kink instabilities, their total pressure needs to be higher than approximately the ram pressure of the infalling gas \citep{gottlieb2022black}. This happens only when the \BH horizon accumulates enough magnetic flux, which powers the jets along with the spin of the \BH, via the BZ process. The high-density gas circulation near the \BH gives rise to magnetic flux polarity flips \citep{Christie2019}, and or this reason, at early times, $t\lesssim 5\times 10^4\rg/c$,  the jets continuously work intermittently, and get disrupted soon after launch.

\begin{figure*}[!t]
\centering
\includegraphics[width=0.7\textwidth]{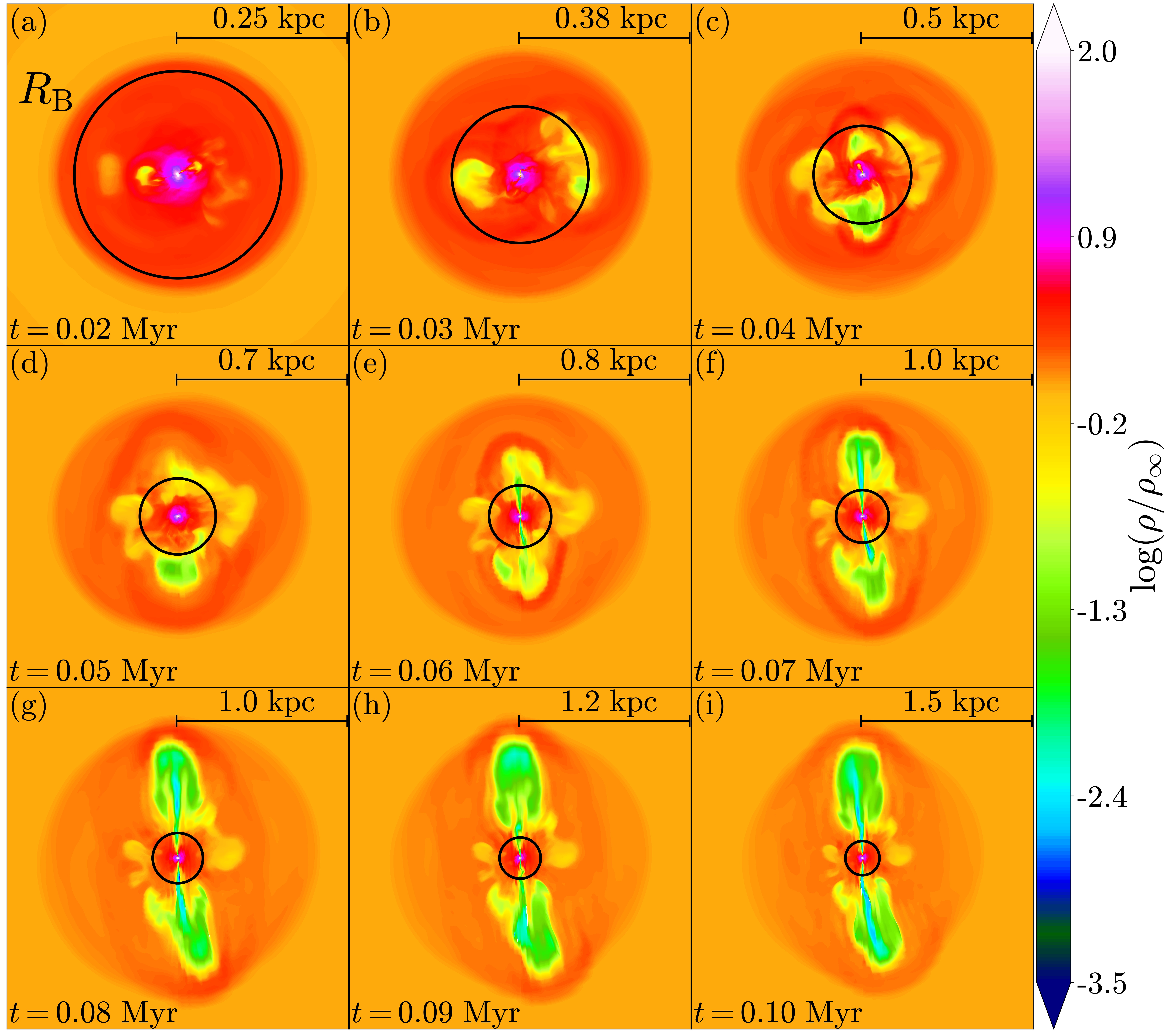}
\caption{The first demonstration that X-shaped radio galaxy morphology naturally emerges from initially axisymmetric conditions: \BH and ambient gas rotating around the vertical $z$-axis, with the gas threaded with a vertical magnetic field.
Panels (a)--(i) show a time sequence of vertical slices through the simulated density (see color bar). The times and lengths shown in each panel have been scaled to M87*: $0.15\ {\rm kpc} = \rb = 10^3 \rg$, $475\ {\rm yr} = \rb/c = 10^3 \rg/c$. The early-time $t\lesssim10^5 R_g/c \sim 0.05$~Myr evolution results in intermittent low-density jets (seen in green) that frequently disrupt due to the onslaught of infalling gas and transient reductions in jet power (see also Fig.~\ref{fig:q_vs_time}). In spite of this, the jets and cavities they inflate manage to reach outside of the Bondi radius, which is shown with a black circle. Throughout the simulation, jets and remnant cavities point in different directions and resemble the X-shaped jet morphology in XRGs. The jets stabilize around $t \sim 10^5R_g/c \sim 0.05$~Myr, once the \BH saturates with the vertical magnetic flux and the disk enters the MAD state: the jets become strong enough to avoid getting deflected by the infalling gas and propagate along the vertical $z$-direction.}
\label{fig:plots_grid}
\end{figure*}

Figure~\ref{fig:q_vs_time}(b) shows that, apart from a few bumps, the magnetic flux steadily increases until $t\lesssim 10^5\rg/c$, and $L/L_{\rm BZ}$ displays several prominent peaks reaching near unity,
Figure~\ref{fig:q_vs_time}(a) shows that the dips in $\dot M$ correlate with the spikes in the magnetic flux, luminosity and efficiency in Fig.~\ref{fig:q_vs_time}(b)--(d). These are the moments when the jets are successfully launched with $ \eta \sim 50\% $ before the turbulent polarity flips disrupts them.

To better understand the behavior of the jets, we determine their tilt $\mathcal T$ and precession $\mathcal P$ angles, by computing the polar and azimuthal centroid positions, respectively, for both the northern and southern jets. For this, we introduce the total-energy to mass-energy flux ratio, $\mu = -u_t (h+\sigma+1)$ \citep{chatterjee2019}, which gives the maximum Lorentz factor the flow could attain if all of its energy converted into kinetic energy. Here $u_t=g_{t\nu}u^{\nu}$ is the covariant time-component of the 4-velocity, $h=(u_g +p_g)/(\rho c^2)$ is the specific enthalpy, and $\sigma \approx 2p_{\rm m}/(\rho c^2)$ is the magnetization. We identify the jets on a sphere of radius $50 \rg$ using the condition $\mu \ge 2$, which selects the relativistic regions (jets) and eliminates non- and mildly-relativistic regions (accretion flow and mildly relativistic outflows).
Surprisingly, Fig.~\ref{fig:q_vs_time}(e)--(f) shows that jet tilt and precession angles vary strongly as they erratically launch. In fact, at early times, $t\lesssim 5\times 10^4\rg/c$, the jets launch nearly perpendicular to the polar axis, essentially in the equatorial plane. We see a clear anti-correlation between the inclination and precession angles (Fig.~\ref{fig:q_vs_time}(e)--(f)) and the energy efficiency of the jets (Fig.~\ref{fig:q_vs_time}(d)): when $\eta$ peaks at $t\sim 7.5\times 10^4\rg/c$, both $\mathcal T$ and $\mathcal P$ drop. Conversely, as $\eta$ drops at $t\sim10^5\rg/c$, both $\mathcal T$ and $\mathcal P$ increase. This suggests that weaker jets are more easily deflected sideways by the pressure of the infalling gas.

Figure~\ref{fig:plots_grid}(a)--(e) shows that the jets, which appear as under-dense green regions, indeed propagate nearly horizontally away from the center. Their launch direction strongly deviates from that of the \BH spin and ambient gas angular momentum vector directions, both of which are vertical and point along the $z$-axis. Shortly after launch, the jets subsequently quench. The timescale between the launching and quenching of the jets is approximately $t\sim 10^4\rg/c$, or $\sim 0.01$ Myr when rescaled to M87*. Furthermore, in a surprising fashion, the intermittent jets substantially wobble in time, both in tilt and precession angles, as the jets struggle to pierce through the dense surrounding gas. 


Later on, at $t\gtrsim 10^5\rg/c$, Fig.~\ref{fig:q_vs_time}(a)--(d) shows that the jet efficiency increases, and all quantities settle around their respective asymptotic values. The accretion rate reaches the low value of $\langle \mdot \rangle/\mdotb=0.02$ in units of the analytic spherical Bondi accretion rate while the magnetic flux asymptotes at the value $\langle \phi_{\rm BH} \rangle  = 50 $. The jet power levels off at $\langle L \rangle = 0.9\ L_{\rm BZ}$ with an outflow efficiency reaching $\langle \eta \rangle=150\%$. Figure~\ref{fig:q_vs_time} shows these steady state average values with horizontal dashed orange lines. Figure~\ref{fig:q_vs_time}(e) shows that the jets manage to avoid getting deflected by the infalling gas and launch with a near-zero tilt, along the vertical $z$-axis. We can also see this in Fig.~\ref{fig:plots_grid}(c)--(i) that shows the jets propagating vertically outwards, displaying only slight bends, producing powerful backflows, and maintaining their overall stability.

The above high values of dimensionless \BH magnetic flux, jet luminosity and efficiency, as well as the ability of the jets to avoid getting deflected by the infalling gas, make it clear that the jets are now strong enough to overcome the destructive effects of the infalling gas and operate at full strength.
In fact, such high jet efficiency values, $\eta\gtrsim 100$\%, are characteristic of the magnetically arrested disk (MAD) state in which the large-scale vertical magnetic flux saturates the \BH and becomes strong enough to periodically overcome the gravity pulling in the accreting gas and escape from the \BH \citep{tchekhovskoy2011efficient}. These repeating magnetic flux eruptions manifest themselves through the fluctuations of the magnetic flux and jet efficiency in Fig.~\ref{fig:q_vs_time}(b) and (d). 
For the rest of the simulation, we do not see any activity that indicates any deviation from this steady-state, although the rest of the duration is as long as the first highly-variable state.

\begin{figure}[htp]
\includegraphics[width=\columnwidth]{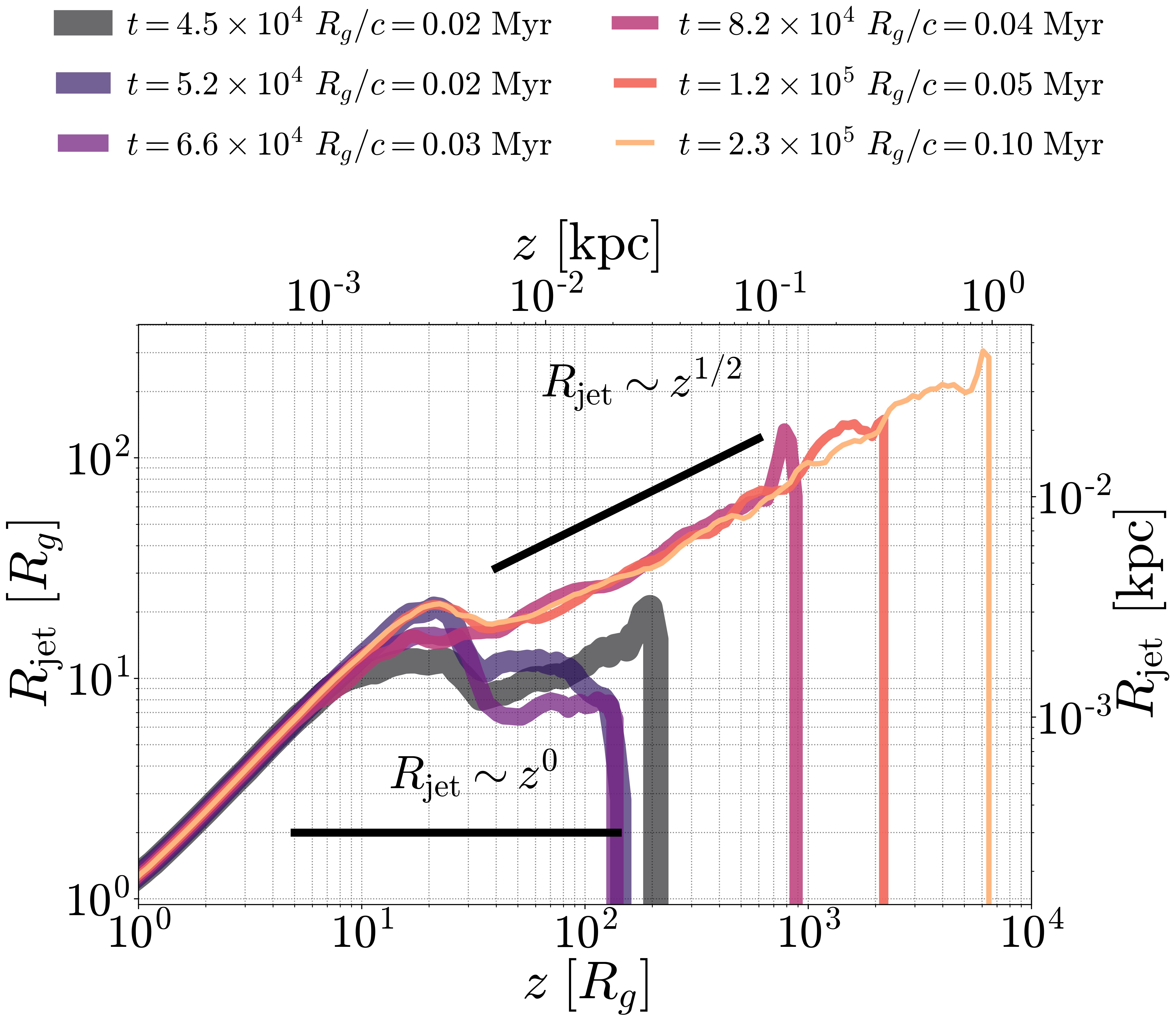}
\caption{Northern jet shape ($R_{\rm jet}$ vs $z$) transitions cylindrical shape at early time to parabolic shape at late time (southern jet shows qualitatively similar behavior; not shown). Different colors and thickness indicate different times (see legend). At $t\lesssim0.03$~Myr, the jets exhibit a cylindrical shape at $z\gtrsim10\rg$. At early times, the ram-pressure is high (large $\dot M$ in Fig.~\ref{fig:q_vs_time}a) and \BH magnetic flux and jet power are low (small $\phi_{\rm BH}$ and $\eta$ in Fig.~\ref{fig:q_vs_time}b,d), leading to strong collimation of the jets. At later times, after enough magnetic flux has accumulated on the \BH, the jets become stronger, and take on a parabolic shape. We compute the cylindrical radius of the jet via the solid angle subtended by the jet (which we identify here as the region of $\mu>1.5$).}%
\label{fig:jet-radius}
\end{figure}

\begin{figure*}[!t]
\centering
\includegraphics[width=0.7\textwidth]{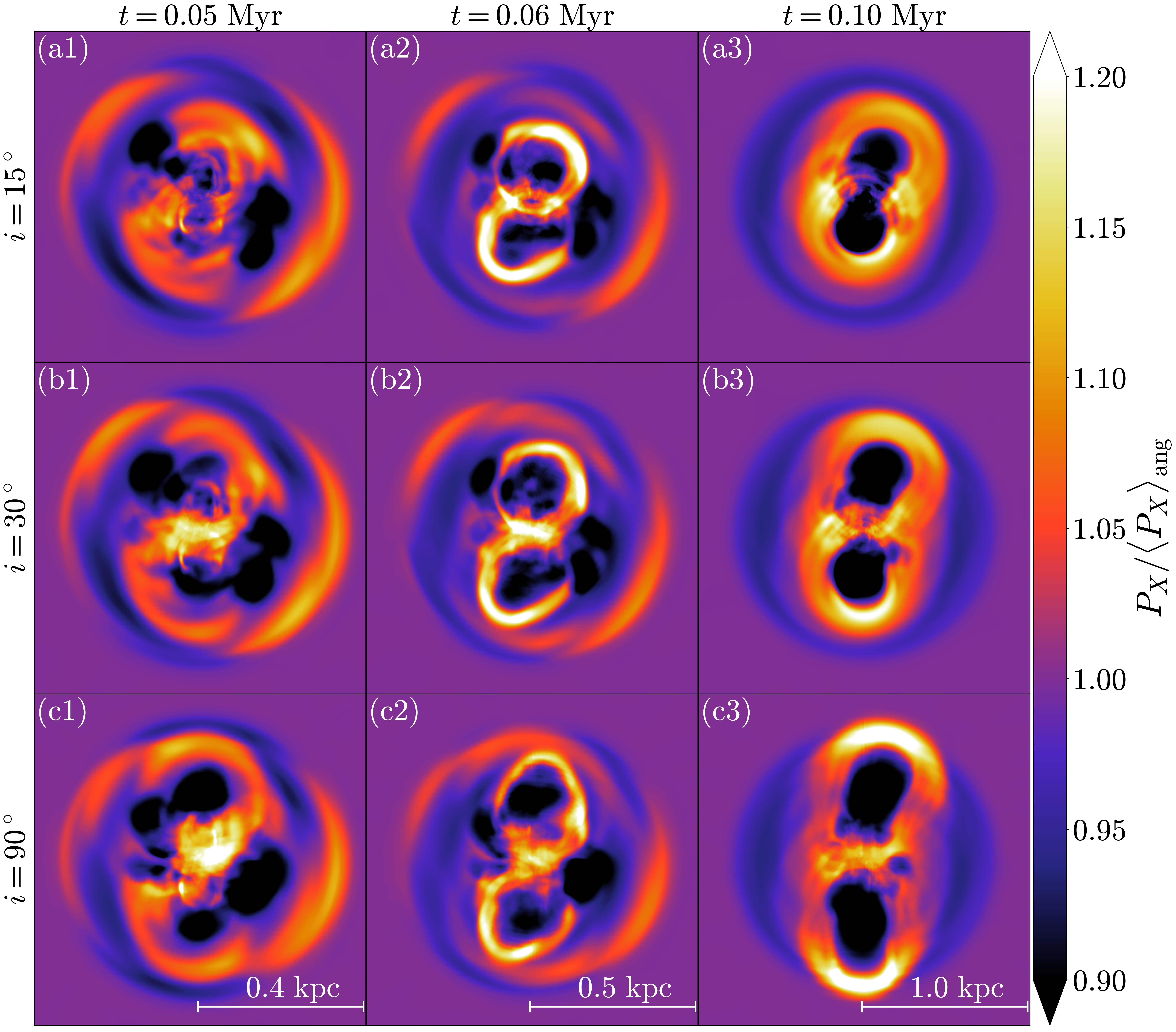}

\caption{Synthetic bolometric X-ray images (emissivity $P_X \propto \rho^2 T^{\!1/2}$, neglecting absorption and scattering) demonstrate the formation of X-shaped jet morphology. Shock compressed and heated regions dominate the Bremsstrahlung emission, whereas empty cavities show jets or jet-inflated cavities. These low-emission regions present a morphology similar to X-shaped jets in XRGs.
The color depicts the emissivity normalized by its angle-averaged value (Sec.~\ref{sec:synthetic}). Each row and column corresponds to a different inclination angle and time, respectively. Times and lengths shown in each panel have been scaled to M87* (see also Fig.~\ref{fig:plots_grid}).
}
\label{fig:xray_synth}
\end{figure*}

The formation of wobbling jets is in contrast with the general expectation that when the \BH\ spin and accreting gas angular momentum vectors are aligned, the accretion disk will produce jets that propagate along their direction. The reason for this discrepancy lies in the interaction of the jets with the surrounding gas, which deflects the jet head. This implies that even though this initial oscillating behavior of the jets might be a transient phenomenon, its effects can be long lasting and potentially probed as lower density cavities in the surrounding medium. The resulting morphology remarkably resembles the strongly asymmetric X-shaped jet morphology of XRGs (Sec.~\ref{sec:intro}), even though the initial set-up is axisymmetric about the $z$-axis, apart from 2\%-level pressure perturbations. Because realistic ISM has much stronger pressure fluctuations, any effect due to wobbling seen here will be much more pronounced for realistic ambient media. The cancellation of gas angular momentum near the \BH can shut-off a jet outburst and power newly-formed pair of jets in a different direction. Eventually, enough magnetic flux accumulation on the \BH leads to a magnetically arrested disk  (MAD, \citealt{tchekhovskoy2011efficient}) and the power of the jet reaches its maximum value and manages to overcome the ambient gas density. In Fig.~\ref{fig:plots_grid}(e)--(i), the system is in its MAD state and not only do the newly formed jets remain stable near the highest density regions around the \BH, but also retain their stability all the way out to $\sim 7000 \rg \simeq 1$~kpc, albeit the jet half-opening angle at these large scales is marginally resolved by about $7$ cells.

\subsection{Cylindrical and Parabolic Jets} \label{sec:jet-shape}

Jets propagating through a dense medium can have rather different shape and stability compared to jets in vacuum. 
Observations indicate that AGN jets often show transition from the parabolic shape inside to conical shape outside the Bondi radius \citep{2013ApJ...775..118N,2020MNRAS.495.3576K, refId0}. The origin of such a transition could either be due to the internal jet evolution \citep[e.g.,][]{2020MNRAS.498.2532N} or due to changes in the ambient density profile \citep[e.g.,][]{barniol_duran2017}.

Figure~\ref{fig:jet-radius} shows the effective cylindrical radius of the northern jet, implied by the solid angle it subtends, versus the distance along the jet (the shape of the southern jet is similar and not shown). Since we use a logarithmic grid, at $r\gtrsim 4000\rg$ numerical diffusion in the jets increases the mixing, and thus we set the condition to isolate the jet to $\mu \ge 1.5$.
In the very early intermittent jet phase, $t\lesssim 7\times 10^4\rg/c = 0.03$~Myr, the jet is approximately cylindrical, $R_{\rm jet} \sim {\rm constant}$. At later times, while still in the early-phase and substantially tilted relative the $z$-axis, the jet transitions to a parabolic shape, $R_{\rm jet} \propto z^{1/2}$. The parabolic shape persists to late times, and the jets stably propagate from the \BH along the $z$-axis out to $z \sim 7000\rg=7\rb\simeq 1$~kpc, without showing any tell-tale signs of disruptive kink-instability. Although the jets leave the Bondi sphere, they do not appear to exhibit the transition from parabolic to conical geometry at the Bondi radius. It is possible that longer and higher resolution simulations will show both the jet disruption, due to the kink instability, and the parabolic-to-conical transition, when the backflows are no longer able to reach from the jet head back to \rb and affect the jet shape there.

\subsection{Shocks and Cavities in Synthetic X-ray Images} \label{sec:synthetic}
Each panel of Fig.~\ref{fig:plots_grid} indicates the position of the Bondi radius with a black circle and the spatial extent of the system with a black scale bar.
Figure~\ref{fig:plots_grid} shows that gas infall launches an expanding spherical accretion shock that forms as the gas finds itself rushing to the \BH too fast. The shock reduces the gas radial infall velocity and helps the system to reach a steady state. Jets drive into the ambient gas an additional pair of bow shocks that by the end of the simulation elongate the accretion shock in the polar directions and start outrunning it.

The accretion and jet bow shocks compress and heat the ambient medium, resulting in thermal Bremsstrahlung X-ray emission. We are presenting a crude estimate of what the X-ray emission would look like, by calculating the Bremsstrahlung bolometric emissivity which scales as $P_X \sim \rho^2 T^{1/2}$ for fully-ionized hydrogen. We assume an optically-thin gas and do not include any absorption or scattering effects.
To construct synthetic X-ray images, we compute the projected fluid variables along the desired line of sight and use them to calculate the emissivity. 

Columns in Fig.~\ref{fig:xray_synth} depict a time series of synthetic X-ray bolometric emissivity images. To highlight structure, we normalize the emissivity by its angle-averaged radial profile.
Rows show views at different inclination angles relative to the $z$-axis, $i = [15,30,90]^{\circ}$. At early times, in both the left and middle columns, both accretion shock and bow shocks are visible, leading to emissivity increase of $\approx 5{-}20\%$ compared to the angle average (see the color bar). 
The bow shock encompasses the bulk of excess X-ray emissivity and surrounds the newly-formed jets. The dark regions show both the old jet-inflated cavities and newly-formed jets. The jet-inflated low-density cavities buoyantly rise, slower than the jets, and are soon outrun by them. The resulting X-shaped morphology is especially prominent in Fig.~\ref{fig:xray_synth}(b1),(b2),(c1),(c2): here, the jets are approximately at the same distance from the \SMBH as the older, jet-inflated cavities, forming a distinct X-shape.
In the right column, even though the bow shock has outrun the accretion shock and the jets have reached $\sim 7\rb \simeq 1$~kpc, we can still see the old jet-inflated remnant cavities closer to the \BH.

\section{Summary and Discussion}\label{sec:discussion}
We have presented the first-ever 3D GRMHD simulation of a \SMBH accreting rotating magnetized ambient gas from outside the Bondi radius, at a scale separation of $\rb/\rg=10^3$ that is unusually large in 3D GRMHD.
Initially located outside an empty cavity of radius \rb, the uniform-density ambient gas contains a weak vertical magnetic field in the $z$-direction, which is aligned with both the \BH spin and gas angular momentum vectors. As the gas falls in, it self-consistently forms a thick accretion torus whose size is set by the circularization radius of the ambient gas (eq.~\ref{eq:rcirc}). 
More generally, the properties of the accretion system are controlled by several manifestly physical dimensionless parameters (see Sec.~\ref{sec:setup}): (i) dimensionless Bondi radius, $\rb/\rg$, (ii) dimensionless circularization radius, $\rcirc/\rg$,  (iii) ambient gas plasma-$\beta$, (iv) dimensionless \BH spin, $a$.
This offers certain advantages over the standard equilibrium torus initial conditions typically used in GRMHD simulations \citep{fis76,2003ApJ...592.1060D}, where the parameters of the torus are harder to relate to observables.

As the infalling gas approaches the sonic surface too fast an accretion shock is developed. Around the same time, $t\simeq 2\times 10^{4}\rg/c$, a low-density intermittent jet launches close to the equatorial plane and disrupts soon thereafter. The disrupted jet remnants forms low-density bubbles, which buoyantly rise beyond the Bondi radius. The repeated disruption and recreation of the jets can be caused by: (i) the highly variable large-scale vertical flux on the \BH horizon (Fig.~\ref{fig:q_vs_time}b), which leads to variability in the jet power (Fig.~\ref{fig:q_vs_time}d) and even intermittent destruction of the jets; (ii) high ram pressure of the infalling gas on the \BH horizon overcomes the ram pressure of the jets at times of low jet power and deflects them towards the equatorial regions. Indeed, minima in the $\phi_{\rm BH}$ and $\eta$ in Fig.~\ref{fig:q_vs_time}(b,d) correlate with the times of jet destruction, $L/L_{\rm BZ}\ll1$, seen in Fig.~\ref{fig:q_vs_time}(c). At early times, when the mass accretion rate is high (large $\mdot/\mdotb\sim0.06$) and jet ram pressure and power are low ($\eta\lesssim50$\%), the tilt and precession angles of our jets reaches large values, $\mathcal{T} \sim 90^{\circ}$ and $\mathcal{P} \sim 180^{\circ}$ (Fig.~\ref{fig:q_vs_time}e,f), implying that the infalling gas has deflected the jets towards the equatorial plane, i.e., perpendicular to the $z$-axis. Figure~\ref{fig:plots_grid} shows that these early-time intermittent equatorial jets inflate buoyant bubbles whose direction is misaligned relative to the late-time stable jets. The synthetic X-ray images in Fig.~\ref{fig:xray_synth} show multiple sets of jet-inflated cavities at different orientations. This morphology strongly resembles X-shaped jets in XRGs. Thus, the deflections of jets by infalling material can provide a natural and simple explanation of XRG formation. Furthermore, the short-lived intermittent jet phase can explain the rarity of XRGs, which make up less than 1 in 10 radio galaxies.

At $t\gtrsim 10^5 \rg/c$, the \BH accumulates so much magnetic flux that the system goes MAD ($\langle \phi_{\rm BH} \rangle = 50$) and the jets attain the maximum power for a given mass accretion rate \citep{tchekhovskoy2011efficient}. 
Namely, the jet luminosity is comparable to the BZ power, $\langle L \rangle \sim L_{\rm BZ}$ (Fig.~\ref{fig:q_vs_time}c), and the outflow energy efficiency reaches $\langle \eta \rangle=150\%$ (Fig.~\ref{fig:q_vs_time}d).
The accretion rate on the \BH saturates at a level of $\langle \mdot \rangle = 0.02 \mdotb$ (Fig.~\ref{fig:q_vs_time}a), implying that $98$\% of the infalling gas at the Bondi radius is ejected from the system, with only $2$\% of the gas reaching the $\BH$.

The Bondi accretion rate for M87* is 
\begin{align}
\mdotb &= 4\pi\lambda_{\rm s} n_{\rm B} m_{\rm p} \rb^2 c (\rb/\rg)^{-1/2} \notag\\
    &\approx 0.12 M_\odot{\rm yr}^{-1} 
\times \frac{n_{\rm B}}{0.17\,{\rm cm}^{-3}} 
\times \left(\frac{\rb}{150\,{\rm pc}}\right)^{3/2},
\end{align}
where $\lambda_{\rm s} = 1/4$ for $\gamma = 5/3$ (\citealt{shapiro_black_holes_1986,di2003accretion}), and we assumed ionized hydrogen. Using this and the steady-state values for the system given above, the simulated mass accretion rate on the \BH scaled to M87* becomes,
\begin{align}
\mdot &= 2.4\times 10^{-3}M_\odot\,{\rm yr^{-1}} 
              \times \frac{\langle\mdot/\mdotb\rangle}{0.02} 
              \times \frac{\mdotb}{0.12M_\odot\,{\rm yr}^{-1}}, 
\intertext{which comes out near the high end of the inferred range for M87*, $\mdot = (3{-}20)\times10^{-4}M_\odot\,{\rm yr}^{-1}$ \citep{akiyama2021first}, and the simulated power of two jets becomes,}
P_{\rm jets} &= 2\times 10^{44}\,{\rm erg\, s}^{-1}
              \times \frac{\langle\eta\rangle}{150\%} 
              \times \frac{\langle\mdot/\mdotb\rangle}{0.02} 
              \times \frac{\mdotb}{0.12M_\odot\,{\rm yr}^{-1}}, 
\end{align}
in agreement with the inferred range of jet power, $10^{42-45}\,{\rm erg\, s}^{-1}$ \citep[e.g.,][]{stawarz2006dynamics,2015ApJ...805..179B,prieto2016central,nemmen2019spin}. Our expectation is that as we increase the scale separation from $\rb/\rg = 10^3$ closer to more physically-motivated values, $10^{5-6}$, more gas will be ejected in outflows, less gas will reach the \BH, and the agreement with the observations will improve.

We measure the shape of the jets to study whether our jets exhibit the parabolic to conical jet shape transition, which is observed near the Bondi radius in some AGN \citep{2013ApJ...775..118N,2020MNRAS.495.3576K, refId0}. Figure~\ref{fig:jet-radius} shows that whereas early-time jets have a cylindrical shape, at later times, $t\simeq 6\times 10^4 \rg/c$, the jets transition to a parabolic shape. We do not see any signs of transition from parabolic to conical. We will investigate this issue in the future, by carrying out higher-resolution and longer-duration simulations, both of which might help to reveal the transition in jet shape (Sec.~\ref{sec:jet-shape}). 

We have only studied a limited parameter space. In the future we will deploy a set of multiple high-resolution simulations and explore the parameter space, with what we think are the most crucial parameters. Specifically: (i) Choose a physically motivated density profile of $\rho \propto r^{-1}$, which is associated with the ISM and ICM. (ii) Vary the scale separation \rb/\rg to lower and higher values, and quantify the differences in accretion rate efficiencies. (iii) Vary the circularization radius \rcirc/\rg, and study the effects the size of the disk has on the accretion rate and wind power. (iv) Reach durations of $t=10^{6-7} \rg/c$ sufficient for the jets to transition into a conical shape, and/or fall victims to instabilities and  get disrupted. (v) Include radiation transport and the effects of cooling on the dynamics of the accretion and power of the outflows.
n introduce feedback in directions the jets normally can't access.

\section*{Acknowledgements}
We thank Claude-Andre Faucher-Giguere for useful comments.
OG is supported by a CIERA Postdoctoral Fellowship.
AT was supported by BSF grant 2020747 and NSF grants
AST-2107839, 
AST-1815304, 
AST-1911080, 
AST-2031997. 
Support for this work was provided by the National Aeronautics and Space Administration through \chandra Award Number TM1-22005X issued by the \chandra X-ray Center, which is operated by the Smithsonian Astrophysical Observatory for and on behalf of the National Aeronautics Space Administration under contract NAS8-03060.
The study of the jet shape (Fig.~\ref{fig:jet-radius}) has
been supported by the Russian Science Foundation, project 20-62-46021.
The authors acknowledge the Texas Advanced Computing Center (TACC) at
The University of Texas at Austin for providing HPC and visualization
resources that have contributed to the research results reported
within this paper via the LRAC allocation AST20011
(\url{http://www.tacc.utexas.edu}). An award of computer time was provided by the Innovative and Novel Computational Impact on Theory and Experiment (INCITE) program under award PHY129. This research used resources of the Oak Ridge Leadership Computing Facility, which is a DOE Office of Science User Facility supported under Contract DE-AC05-00OR22725. IZ is partially supported by a Clare Boothe Luce Professorship from the Henry Luce Foundation.

%

\bibliography{ms.bbl}{}

\begin{thebibliography}{}
\expandafter\ifx\csname natexlab\endcsname\relax\def\natexlab#1{#1}\fi
\providecommand{\url}[1]{\href{#1}{#1}}
\providecommand{\dodoi}[1]{doi:~\href{http://doi.org/#1}{\nolinkurl{#1}}}
\providecommand{\doeprint}[1]{\href{http://ascl.net/#1}{\nolinkurl{http://ascl.net/#1}}}
\providecommand{\doarXiv}[1]{\href{https://arxiv.org/abs/#1}{\nolinkurl{https://arxiv.org/abs/#1}}}

\bibitem[{Akiyama {et~al.}(2021)Akiyama, Algaba, Alberdi, Alef, Anantua, Asada,
  Azulay, Baczko, Ball, Balokovi{\'c}, {et~al.}}]{akiyama2021first}
Akiyama, K., Algaba, J.~C., Alberdi, A., {et~al.} 2021, The Astrophysical
  Journal Letters, 910, L13

\bibitem[{{Angl{\'e}s-Alc{\'a}zar} {et~al.}(2017){Angl{\'e}s-Alc{\'a}zar},
  {Faucher-Gigu{\`e}re}, {Quataert}, {Hopkins}, {Feldmann}, {Torrey}, {Wetzel},
  \& {Kere{\v{s}}}}]{alcazar2017}
{Angl{\'e}s-Alc{\'a}zar}, D., {Faucher-Gigu{\`e}re}, C.-A., {Quataert}, E.,
  {et~al.} 2017, \mnras, 472, L109, \dodoi{10.1093/mnrasl/slx161}

\bibitem[{{Angl{\'e}s-Alc{\'a}zar} {et~al.}(2015){Angl{\'e}s-Alc{\'a}zar},
  {{\"O}zel}, {Dav{\'e}}, {Katz}, {Kollmeier}, \& {Oppenheimer}}]{alcazar2015}
{Angl{\'e}s-Alc{\'a}zar}, D., {{\"O}zel}, F., {Dav{\'e}}, R., {et~al.} 2015,
  \apj, 800, 127, \dodoi{10.1088/0004-637X/800/2/127}

\bibitem[{{Angl{\'e}s-Alc{\'a}zar} {et~al.}(2021){Angl{\'e}s-Alc{\'a}zar},
  {Quataert}, {Hopkins}, {Somerville}, {Hayward}, {Faucher-Gigu{\`e}re},
  {Bryan}, {Kere{\v{s}}}, {Hernquist}, \& {Stone}}]{2021ApJ...917...53A}
{Angl{\'e}s-Alc{\'a}zar}, D., {Quataert}, E., {Hopkins}, P.~F., {et~al.} 2021,
  \apj, 917, 53, \dodoi{10.3847/1538-4357/ac09e8}

\bibitem[{{Barniol Duran} {et~al.}(2017){Barniol Duran}, {Tchekhovskoy}, \&
  {Giannios}}]{barniol_duran2017}
{Barniol Duran}, R., {Tchekhovskoy}, A., \& {Giannios}, D. 2017, \mnras, 469,
  4957, \dodoi{10.1093/mnras/stx1165}

\bibitem[{{Blandford} \& {Znajek}(1977)}]{blanford1977}
{Blandford}, R.~D., \& {Znajek}, R.~L. 1977, \mnras, 179, 433,
  \dodoi{10.1093/mnras/179.3.433}

\bibitem[{{Boccardi} {et~al.}(2021){Boccardi}, {Perucho, M.}, {Casadio, C.},
  {Grandi, P.}, {Macconi, D.}, {Torresi, E.}, {Pellegrini, S.}, {Krichbaum, T.
  P.}, {Kadler, M.}, {Giovannini, G.}, {Karamanavis, V.}, {Ricci, L.}, {Madika,
  E.}, {Bach, U.}, {Ros, E.}, {Giroletti, M.}, \& {Zensus, J. A.}}]{refId0}
{Boccardi}, B., {Perucho, M.}, {Casadio, C.}, {et~al.} 2021, A\&A, 647, A67,
  \dodoi{10.1051/0004-6361/202039612}

\bibitem[{{Bondi}(1952)}]{bondi1952}
{Bondi}, H. 1952, \mnras, 112, 195, \dodoi{10.1093/mnras/112.2.195}

\bibitem[{{Broderick} {et~al.}(2015){Broderick}, {Narayan}, {Kormendy},
  {Perlman}, {Rieke}, \& {Doeleman}}]{2015ApJ...805..179B}
{Broderick}, A.~E., {Narayan}, R., {Kormendy}, J., {et~al.} 2015, \apj, 805,
  179, \dodoi{10.1088/0004-637X/805/2/179}

\bibitem[{Bruni {et~al.}(2019)Bruni, Panessa, Bassani, Chiaraluce, Kraus,
  Dallacasa, Bazzano, Hern{\'a}ndez-Garc{\'\i}a, Malizia, Ubertini,
  {et~al.}}]{bruni2019discovery}
Bruni, G., Panessa, F., Bassani, L., {et~al.} 2019, The Astrophysical Journal,
  875, 88

\bibitem[{Bruni {et~al.}(2020)Bruni, Panessa, Bassani, Dallacasa, Venturi,
  Saripalli, Brienza, Hern{\'a}ndez-Garc{\'\i}a, Chiaraluce, Ursini,
  {et~al.}}]{bruni2020hard}
---. 2020, Monthly Notices of the Royal Astronomical Society, 494, 902

\bibitem[{{Chatterjee} {et~al.}(2019){Chatterjee}, {Liska}, {Tchekhovskoy}, \&
  {Markoff}}]{chatterjee2019}
{Chatterjee}, K., {Liska}, M., {Tchekhovskoy}, A., \& {Markoff}, S.~B. 2019,
  \mnras, 490, 2200, \dodoi{10.1093/mnras/stz2626}

\bibitem[{{Christie} {et~al.}(2019){Christie}, {Lalakos}, {Tchekhovskoy},
  {Fern{\'a}ndez}, {Foucart}, {Quataert}, \& {Kasen}}]{Christie2019}
{Christie}, I.~M., {Lalakos}, A., {Tchekhovskoy}, A., {et~al.} 2019, \mnras,
  490, 4811, \dodoi{10.1093/mnras/stz2552}

\bibitem[{Cotton {et~al.}(2020)Cotton, Thorat, Condon, Frank, J{\'o}zsa, White,
  Deane, Oozeer, Atemkeng, Bester, {et~al.}}]{cotton2020hydrodynamical}
Cotton, W., Thorat, K., Condon, J., {et~al.} 2020, Monthly Notices of the Royal
  Astronomical Society, 495, 1271

\bibitem[{{Cunningham} {et~al.}(2012){Cunningham}, {McKee}, {Klein},
  {Krumholz}, \& {Teyssier}}]{2012ApJ...744..185C}
{Cunningham}, A.~J., {McKee}, C.~F., {Klein}, R.~I., {Krumholz}, M.~R., \&
  {Teyssier}, R. 2012, \apj, 744, 185, \dodoi{10.1088/0004-637X/744/2/185}

\bibitem[{{De Villiers} \& {Hawley}(2003)}]{2003ApJ...592.1060D}
{De Villiers}, J.-P., \& {Hawley}, J.~F. 2003, \apj, 592, 1060,
  \dodoi{10.1086/375866}

\bibitem[{Di~Matteo {et~al.}(2003)Di~Matteo, Allen, Fabian, Wilson, \&
  Young}]{di2003accretion}
Di~Matteo, T., Allen, S.~W., Fabian, A.~C., Wilson, A.~S., \& Young, A.~J.
  2003, The Astrophysical Journal, 582, 133

\bibitem[{{Fishbone} \& {Moncrief}(1976)}]{fis76}
{Fishbone}, L.~G., \& {Moncrief}, V. 1976, \apj, 207, 962

\bibitem[{{Forman} {et~al.}(2017){Forman}, {Churazov}, {Jones}, {Heinz},
  {Kraft}, \& {Vikhlinin}}]{forman2017}
{Forman}, W., {Churazov}, E., {Jones}, C., {et~al.} 2017, \apj, 844, 122,
  \dodoi{10.3847/1538-4357/aa70e4}

\bibitem[{Gottlieb {et~al.}(2022)Gottlieb, Lalakos, Bromberg, Liska, \&
  Tchekhovskoy}]{gottlieb2022black}
Gottlieb, O., Lalakos, A., Bromberg, O., Liska, M., \& Tchekhovskoy, A. 2022,
  Monthly Notices of the Royal Astronomical Society, 510, 4962

\bibitem[{Jia {et~al.}(2022)Jia, White, Quataert, \&
  Ressler}]{jia2022observational}
Jia, H., White, C.~J., Quataert, E., \& Ressler, S.~M. 2022, arXiv preprint
  arXiv:2201.08431

\bibitem[{{Kaaz} {et~al.}(2019){Kaaz}, {Antoni}, \& {Ramirez-Ruiz}}]{kaaz2019}
{Kaaz}, N., {Antoni}, A., \& {Ramirez-Ruiz}, E. 2019, \apj, 876, 142,
  \dodoi{10.3847/1538-4357/ab158b}

\bibitem[{{Kaaz} {et~al.}(2022){Kaaz}, {Murguia-Berthier}, {Chatterjee},
  {Liska}, \& {Tchekhovskoy}}]{2022arXiv220111753K}
{Kaaz}, N., {Murguia-Berthier}, A., {Chatterjee}, K., {Liska}, M., \&
  {Tchekhovskoy}, A. 2022, arXiv e-prints, arXiv:2201.11753.
\newblock \doarXiv{2201.11753}

\bibitem[{{Kovalev} {et~al.}(2020){Kovalev}, {Pushkarev}, {Nokhrina}, {Plavin},
  {Beskin}, {Chernoglazov}, {Lister}, \& {Savolainen}}]{2020MNRAS.495.3576K}
{Kovalev}, Y.~Y., {Pushkarev}, A.~B., {Nokhrina}, E.~E., {et~al.} 2020, \mnras,
  495, 3576, \dodoi{10.1093/mnras/staa1121}

\bibitem[{{Lal} \& {Rao}(2007)}]{2007MNRAS.374.1085L}
{Lal}, D.~V., \& {Rao}, A.~P. 2007, \mnras, 374, 1085,
  \dodoi{10.1111/j.1365-2966.2006.11225.x}

\bibitem[{{Leahy} \& {Williams}(1984)}]{1984MNRAS.210..929L}
{Leahy}, J.~P., \& {Williams}, A.~G. 1984, \mnras, 210, 929,
  \dodoi{10.1093/mnras/210.4.929}

\bibitem[{{Li} {et~al.}(2013){Li}, {Ostriker}, \& {Sunyaev}}]{li2013}
{Li}, J., {Ostriker}, J., \& {Sunyaev}, R. 2013, \apj, 767, 105,
  \dodoi{10.1088/0004-637X/767/2/105}

\bibitem[{{Li} {et~al.}(2020){Li}, {Gendron-Marsolais}, {Zhuravleva}, {Xu},
  {Simionescu}, {Tremblay}, {Lochhaas}, {Bryan}, {Quataert}, {Murray},
  {Boselli}, {Hlavacek-Larrondo}, {Zheng}, {Fossati}, {Li}, {Emsellem},
  {Sarzi}, {Arzamasskiy}, \& {Vishniac}}]{2020ApJ...889L...1L}
{Li}, Y., {Gendron-Marsolais}, M.-L., {Zhuravleva}, I., {et~al.} 2020, \apjl,
  889, L1, \dodoi{10.3847/2041-8213/ab65c7}

\bibitem[{Liska {et~al.}(2018)Liska, Hesp, Tchekhovskoy, Ingram, van~der Klis,
  \& Markoff}]{liska2018formation}
Liska, M., Hesp, C., Tchekhovskoy, A., {et~al.} 2018, Monthly Notices of the
  Royal Astronomical Society: Letters, 474, L81

\bibitem[{{Liska} {et~al.}(2019){Liska}, {Chatterjee}, {Tchekhovskoy}, {Yoon},
  {van Eijnatten}, {Hesp}, {Markoff}, {Ingram}, \& {van der
  Klis}}]{2019arXiv191210192L}
{Liska}, M., {Chatterjee}, K., {Tchekhovskoy}, A.~e., {et~al.} 2019, arXiv
  e-prints, arXiv:1912.10192.
\newblock \doarXiv{1912.10192}

\bibitem[{Martizzi {et~al.}(2019)Martizzi, Quataert, Faucher-Gigu{\`e}re, \&
  Fielding}]{martizzi2019simulations}
Martizzi, D., Quataert, E., Faucher-Gigu{\`e}re, C.-A., \& Fielding, D. 2019,
  Monthly Notices of the Royal Astronomical Society, 483, 2465

\bibitem[{{McKinney} {et~al.}(2013){McKinney}, {Tchekhovskoy}, \&
  {Blandford}}]{2013Sci...339...49M}
{McKinney}, J.~C., {Tchekhovskoy}, A., \& {Blandford}, R.~D. 2013, Science,
  339, 49, \dodoi{10.1126/science.1230811}

\bibitem[{{Merritt} \& {Ekers}(2002)}]{2002Sci...297.1310M}
{Merritt}, D., \& {Ekers}, R.~D. 2002, Science, 297, 1310,
  \dodoi{10.1126/science.1074688}

\bibitem[{{Nakamura} \& {Asada}(2013)}]{2013ApJ...775..118N}
{Nakamura}, M., \& {Asada}, K. 2013, \apj, 775, 118,
  \dodoi{10.1088/0004-637X/775/2/118}

\bibitem[{Nemmen(2019)}]{nemmen2019spin}
Nemmen, R. 2019, The Astrophysical Journal Letters, 880, L26

\bibitem[{{Nokhrina} {et~al.}(2020){Nokhrina}, {Kovalev}, \&
  {Pushkarev}}]{2020MNRAS.498.2532N}
{Nokhrina}, E.~E., {Kovalev}, Y.~Y., \& {Pushkarev}, A.~B. 2020, \mnras, 498,
  2532, \dodoi{10.1093/mnras/staa2458}

\bibitem[{{Palit} {et~al.}(2019){Palit}, {Janiuk}, \& {Sukova}}]{palit2019time}
{Palit}, I., {Janiuk}, A., \& {Sukova}, P. 2019, \mnras, 487, 755,
  \dodoi{10.1093/mnras/stz1296}

\bibitem[{Parma {et~al.}(1985)Parma, Ekers, \& Fanti}]{parma1985high}
Parma, P., Ekers, R., \& Fanti, R. 1985, Astronomy and Astrophysics Supplement
  Series, 59, 511

\bibitem[{{Porth} {et~al.}(2019){Porth}, {Chatterjee}, {Narayan}, {Gammie},
  {Mizuno}, {Anninos}, {Baker}, {Bugli}, {Chan}, {Davelaar}, {Del Zanna},
  {Etienne}, {Fragile}, {Kelly}, {Liska}, {Markoff}, {McKinney}, {Mishra},
  {Noble}, {Olivares}, {Prather}, {Rezzolla}, {Ryan}, {Stone}, {Tomei},
  {White}, {Younsi}, {Akiyama}, {Alberdi}, {Alef}, {Asada}, {Azulay}, {Baczko},
  {Ball}, {Balokovi{\'c}}, {Barrett}, {Bintley}, {Blackburn}, {Boland},
  {Bouman}, {Bower}, {Bremer}, {Brinkerink}, {Brissenden}, {Britzen},
  {Broderick}, {Broguiere}, {Bronzwaer}, {Byun}, {Carlstrom}, {Chael},
  {Chatterjee}, {Chen}, {Chen}, {Cho}, {Christian}, {Conway}, {Cordes},
  {Geoffrey}, {Crew}, {Cui}, {De Laurentis}, {Deane}, {Dempsey}, {Desvignes},
  {Doeleman}, {Eatough}, {Falcke}, {Fish}, {Fomalont}, {Fraga-Encinas},
  {Freeman}, {Friberg}, {Fromm}, {G{\'o}mez}, {Galison}, {Garc{\'\i}a},
  {Gentaz}, {Georgiev}, {Goddi}, {Gold}, {Gu}, {Gurwell}, {Hada}, {Hecht},
  {Hesper}, {Ho}, {Ho}, {Honma}, {Huang}, {Huang}, {Hughes}, {Ikeda}, {Inoue},
  {Issaoun}, {James}, {Jannuzi}, {Janssen}, {Jeter}, {Jiang}, {Johnson},
  {Jorstad}, {Jung}, {Karami}, {Karuppusamy}, {Kawashima}, {Keating},
  {Kettenis}, {Kim}, {Kim}, {Kim}, {Kino}, {Koay}, {Patrick}, {Koch}, {Koyama},
  {Kramer}, {Kramer}, {Krichbaum}, {Kuo}, {Lauer}, {Lee}, {Li}, {Li},
  {Lindqvist}, {Liu}, {Liuzzo}, {Lo}, {Lobanov}, {Loinard}, {Lonsdale}, {Lu},
  {MacDonald}, {Mao}, {Marrone}, {Marscher}, {Mart{\'\i}-Vidal}, {Matsushita},
  {Matthews}, {Medeiros}, {Menten}, {Mizuno}, {Moran}, {Moriyama},
  {Moscibrodzka}, {M{\"u}ller}, {Nagai}, {Nagar}, {Nakamura}, {Narayanan},
  {Natarajan}, {Neri}, {Ni}, {Noutsos}, {Okino}, {Oyama}, {{\"O}zel},
  {Palumbo}, {Patel}, {Pen}, {Pesce}, {Pi{\'e}tu}, {Plambeck}, {PopStefanija},
  {Preciado-L{\'o}pez}, {Psaltis}, {Pu}, {Ramakrishnan}, {Rao}, {Rawlings},
  {Raymond}, {Ripperda}, {Roelofs}, {Rogers}, {Ros}, {Rose}, {Roshanineshat},
  {Rottmann}, {Roy}, {Ruszczyk}, {Rygl}, {S{\'a}nchez},
  {S{\'a}nchez-Arguelles}, {Sasada}, {Savolainen}, {Schloerb}, {Schuster},
  {Shao}, {Shen}, {Small}, {Sohn}, {SooHoo}, {Tazaki}, {Tiede}, {Tilanus},
  {Titus}, {Toma}, {Torne}, {Trent}, {Trippe}, {Tsuda}, {van Bemmel}, {van
  Langevelde}, {van Rossum}, {Wagner}, {Wardle}, {Weintroub}, {Wex}, {Wharton},
  {Wielgus}, {Wong}, {Wu}, {Young}, {Young}, {Yuan}, {Yuan}, {Zensus}, {Zhao},
  {Zhao}, {Zhu}, \& {Event Horizon Telescope Collaboration}}]{Porth:19}
{Porth}, O., {Chatterjee}, K., {Narayan}, R., {et~al.} 2019, \apjs, 243, 26,
  \dodoi{10.3847/1538-4365/ab29fd}

\bibitem[{Prieto {et~al.}(2016)Prieto, Fern{\'a}ndez-Ontiveros, Markoff,
  Espada, \& Gonz{\'a}lez-Mart{\'\i}n}]{prieto2016central}
Prieto, M., Fern{\'a}ndez-Ontiveros, J., Markoff, S., Espada, D., \&
  Gonz{\'a}lez-Mart{\'\i}n, O. 2016, Monthly Notices of the Royal Astronomical
  Society, 457, 3801

\bibitem[{{Proga} \& {Begelman}(2003{\natexlab{a}})}]{2003ApJ...582...69P}
{Proga}, D., \& {Begelman}, M.~C. 2003{\natexlab{a}}, \apj, 582, 69,
  \dodoi{10.1086/344537}

\bibitem[{{Proga} \& {Begelman}(2003{\natexlab{b}})}]{2003ApJ...592..767P}
---. 2003{\natexlab{b}}, \apj, 592, 767, \dodoi{10.1086/375773}

\bibitem[{Ressler {et~al.}(2021)Ressler, Quataert, White, \&
  Blaes}]{ressler2021magnetically}
Ressler, S.~M., Quataert, E., White, C.~J., \& Blaes, O. 2021, Monthly Notices
  of the Royal Astronomical Society, 504, 6076

\bibitem[{Roberts {et~al.}(2015)Roberts, Saripalli, \&
  Subrahmanyan}]{roberts2015abundance}
Roberts, D.~H., Saripalli, L., \& Subrahmanyan, R. 2015, The Astrophysical
  Journal Letters, 810, L6

\bibitem[{{Russell} {et~al.}(2015){Russell}, {Fabian}, {McNamara}, \&
  {Broderick}}]{russell2015}
{Russell}, H.~R., {Fabian}, A.~C., {McNamara}, B.~R., \& {Broderick}, A.~E.
  2015, \mnras, 451, 588, \dodoi{10.1093/mnras/stv954}

\bibitem[{{Shapiro} \& {Teukolsky}(1986)}]{shapiro_black_holes_1986}
{Shapiro}, S.~L., \& {Teukolsky}, S.~A. 1986, {Black Holes, White Dwarfs and
  Neutron Stars: The Physics of Compact Objects} (Black Holes, White Dwarfs and
  Neutron Stars: The Physics of Compact Objects, by Stuart L.~Shapiro, Saul
  A.~Teukolsky, pp.~672.~ISBN 0-471-87316-0.~Wiley-VCH , June 1986.)

\bibitem[{Stawarz {et~al.}(2006)Stawarz, Aharonian, Kataoka, Ostrowski,
  Siemiginowska, \& Sikora}]{stawarz2006dynamics}
Stawarz, {\L}., Aharonian, F., Kataoka, J., {et~al.} 2006, Monthly Notices of
  the Royal Astronomical Society, 370, 981

\bibitem[{{Sukov{\'a}} {et~al.}(2017){Sukov{\'a}}, {Charzy{\'n}ski}, \&
  {Janiuk}}]{2017MNRAS.472.4327S}
{Sukov{\'a}}, P., {Charzy{\'n}ski}, S., \& {Janiuk}, A. 2017, \mnras, 472,
  4327, \dodoi{10.1093/mnras/stx2254}

\bibitem[{{Sukov{\'a}} \& {Janiuk}(2015)}]{2015MNRAS.447.1565S}
{Sukov{\'a}}, P., \& {Janiuk}, A. 2015, \mnras, 447, 1565,
  \dodoi{10.1093/mnras/stu2544}

\bibitem[{{Tchekhovskoy} \& {Bromberg}(2016)}]{tchekhovskoy2016}
{Tchekhovskoy}, A., \& {Bromberg}, O. 2016, \mnras, 461, L46,
  \dodoi{10.1093/mnrasl/slw064}

\bibitem[{Tchekhovskoy {et~al.}(2011)Tchekhovskoy, Narayan, \&
  McKinney}]{tchekhovskoy2011efficient}
Tchekhovskoy, A., Narayan, R., \& McKinney, J.~C. 2011, Monthly Notices of the
  Royal Astronomical Society: Letters, 418, L79

\bibitem[{{Waters} {et~al.}(2020){Waters}, {Aykutalp}, {Proga}, {Johnson},
  {Li}, \& {Smidt}}]{waters2020}
{Waters}, T., {Aykutalp}, A., {Proga}, D., {et~al.} 2020, \mnras, 491, L76,
  \dodoi{10.1093/mnrasl/slz168}

\bibitem[{{Xu} \& {Stone}(2019)}]{2019MNRAS.488.5162X}
{Xu}, W., \& {Stone}, J.~M. 2019, \mnras, 488, 5162,
  \dodoi{10.1093/mnras/stz2002}

\bibitem[{{Yang} {et~al.}(2019){Yang}, {Joshi}, {Gopal-Krishna}, {An}, {Ho},
  {Wiita}, {Liu}, {Yang}, {Wang}, {Wu}, \& {Yang}}]{2019ApJS..245...17Y}
{Yang}, X., {Joshi}, R., {Gopal-Krishna}, {et~al.} 2019, \apjs, 245, 17,
  \dodoi{10.3847/1538-4365/ab4811}

\bibitem[{{Zhuravleva} {et~al.}(2016){Zhuravleva}, {Churazov}, {Ar{\'e}valo},
  {Schekochihin}, {Forman}, {Allen}, {Simionescu}, {Sunyaev}, {Vikhlinin}, \&
  {Werner}}]{2016MNRAS.458.2902Z}
{Zhuravleva}, I., {Churazov}, E., {Ar{\'e}valo}, P., {et~al.} 2016, \mnras,
  458, 2902, \dodoi{10.1093/mnras/stw520}

\end{thebibliography}
\bibliographystyle{aasjournal}



\end{document}